# Dissipative instability of shock waves

Sergey G. Chefranov

Obukhov Institute of Atmospheric Physics, Russian Academy of Sciences,

Moscow, Russia

Physics Department, Technion-Israel Institute of Technology, Haifa 32000, Israel

schefranov@mail.ru ; csergei@technion.ac.il

## Abstract

A new condition for the linear dissipative instability of the strong plane shock wave in an arbitrary medium is obtained. The instability of the shock is realized due to the flow instability behind its front, which is similar to the known dissipative instability in the boundary layer for the Tollmien-Schlichting waves. It is found that within the limit of low viscosity the one-dimensional longitudinal disturbances grow much faster than the two-dimensional corrugation ones. It points to a better correspondence to experiment of the new condition for the absolute instability of the shock in comparison with the S.P. D'yakov classical theory (1954), which does not take viscosity into account.

## 1. Introduction

Shock waves arise in many nonlinear processes in plasma physics, hydrodynamics, aerodynamics and astrophysics [1-7]. So, the existence of the unresolved stability problem for shocks is very important challenge up to now. Indeed, the known classical linear theory of shock waves stability in ideal inviscid media [8–10] does not explain the instability of the plane shock waves observed in real gases [11]. Moreover, in [12] on the base of the Burgers equation, for weak shocks with finite width due to viscosity [13], the stability to any one-dimensional (and also to the two-dimensional [14]) disturbances is also obtained in contradiction with experiment [11].

For strong shocks in the D'yakov theory [8] the instability conditions are obtained (see below (1.1) and (1.2)), but only for the transverse two-dimensional corrugation disturbances. In [8], as in the theory [10, 12], the plane shocks are always stable with respect to the longitudinal (along a direction of the shock velocity) one-dimensional disturbances.

The D'yakov theory generalization is given in [15], where the viscosity is taken into account already for the case of strong shocks. In [15] the destabilizing role of viscosity is identified only for the mode, which in theory [8-10] corresponds to the neutrally stable (with constant amplitude in time) acoustic mode. However, in [15] any changes in the conditions of absolute instability of the theory [8] are not obtained. So in [15] and in [8] for two-dimensional disturbances the conditions of absolute instability are represented in the same form:

$$h < -1 \tag{1.1}$$

$$h > 1 + 2M \tag{1.2}$$

In (1.2) $M < 1$ is a Mach number in the compression region behind the shock wave front. In (1.1) and (1.2) $h = j^2 (\frac{dV}{dp})_H$ is the dimensionless "D'yakov parameter" [8] representing the ratio of Rayleigh line and the Hugoniot curve slopes, where $j$ is the mass flux density across the shock [10]. At the same time, for the case of one-dimensional longitudinal disturbances in [15] the plane shock wave is always stable, as in [8-10, 12].

In this paper the instability of shocks only to the acoustic disturbances are considered for the first time, in contrast to [8-10, 15] where the entropy disturbances are need to consider in addition. Indeed, in [8-10] the instability of shocks is impossible at all if entropy disturbances are not taken into account. Also for the first time here it is stated the possibility of instability of shock with respect to the longitudinal one-dimensional disturbances. Moreover, in the limit of low viscosity one-dimensional disturbances grow faster than two-dimensional corrugation perturbations. In this case the increment of exponentially growing of one-dimensional disturbances is proportional to parameter $\tau^{-1} = 2w^2/\nu$, which also was introduced in [12], where $|w|$ is the jump of velocity at the shock discontinuity, and $\nu$ is the kinematic viscosity.

As a main result, a dissipative mechanism of flow instability in compression region behind the plane shock wave front is stated. This instability of shock is similar to the one that is realized for flows in the boundary layers at sufficiently high Reynolds numbers due to finite viscosity [10, 16]. But in this case, there is a specific feature of the resulting shock instability, which distinguishes it from the usual instability in the boundary layer, when Tollmien-Schlichting waves are arising in it. Indeed, instability occurs first in the flow in the compression region behind the plane shock wave front, where the Mach number $M < 1$. Thus, the growing perturbations are not just carried out by the flow, as is the case in the boundary layer, but they spread at the speed of sound and catching up with the shock wave front and perturb it so that it is destroyed after some time.

Also a new condition of absolute instability of shocks is obtained when for one-dimensional and for two-dimensional disturbances it has the same form:

$$1 - 2M^2 < h < 1 \tag{1.3}$$

The condition of instability (1.3) at some not very small Mach's number $0 < M < 1$, in contrast to (1.1) and (1.2) already does not contradict to the experimental data [11], because in [11] the instability of the plane shock wave front in argon and carbon dioxide is observed for the values of the D'yakov parameter near $h \approx 0.05$.

Concretization for $h$ and condition (1.3) herein are carried out for shock waves in fluids and gases through the use of known from experiment representations for the Huganiot curve (in the form of a linear dependence of the shock wave velocity on medium velocity behind the front [17-19]). In this connection, the results [7, 20-22] of numerical simulation of the shock wave instability are also considered.

## 2. The derivation of the characteristic equation.

1. Let us consider a plane stationary shock wave of arbitrary intensity propagating in an arbitrary medium in the positive direction of the axis $x$.

Afterwards, we use the same notations as in [9], introducing the subscript 0 to denote the values related to the region in before the shock wave front in the region $x > 0$ in the coordinate system in which the plane front of the shock wave is motionless and located at point $x = 0$.

We consider only disturbances related to the compression region $x < 0$. For region $x > 0$, the medium remains undisturbed, since, as usual, the condition $M_0 > 0$ on the Mach number $M_0 = |w_0|/c_0; w_0 = U_0 - D < 0$ in the undisturbed medium is considered [8-10]. Besides, $D, U_0, c_0$ are velocities of the shock wave front, the undisturbed medium particles and sound velocity in medium for $x > 0$. The velocity of the undisturbed medium is usually taken equal to zero $U_0 = 0$. The Mach number in the compression region $M = |w|/c; w = U - D < 0$ is also determined in a similar way [8-10]. Besides, $U; c$ are velocities of the disturbed medium particles following the shock wave front and the local speed of sound. As usual, condition $M < 1$ [8-10] is considered.

For simplicity, we consider only the case, when the shock front is assumed to be uniform in the direction of the $y$ axis located in plane $(z, y)$ and there are no disturbances of the velocity field with a component along this axis. Moreover, the equation for the disturbed plane surface of the shock wave has the form corresponding to the dependence on the coordinate $z$ for the amplitude of deviations from the equilibrium position $x = 0$ along the axis $x$ [9]:

$$x_{sh} = g(z,t) \qquad (2.1)$$

In the coordinate system in which the undisturbed front of the shock wave is stationary, the equations for small disturbances of the velocity and pressure fields behind the front of the shock wave in the region $x<0$ in the linear approximation are the following [15] (when in [15] $v_2=0$):

$$\frac{\partial V_{1z}}{\partial t}+w\frac{\partial V_{1z}}{\partial x}=-\frac{1}{\rho}\frac{\partial p_1}{\partial z}+v_1\Delta_2 V_{1z}+(\frac{v_1}{3}+v_2)\frac{\partial div\vec{V_1}}{\partial z};\Delta_2=\frac{\partial^2}{\partial x^2}+\frac{\partial^2}{\partial z^2}$$

$$\frac{\partial V_{1x}}{\partial t}+w\frac{\partial V_{1x}}{\partial x}=-\frac{1}{\rho}\frac{\partial p_1}{\partial x}+v_1\Delta_2 V_{1x}+(\frac{v_1}{3}+v_2)\frac{\partial div\vec{V_1}}{\partial x};div\vec{V_1}=\frac{\partial V_{1z}}{\partial z}+\frac{\partial V_{1x}}{\partial x} \qquad (2.2)$$

$$\frac{\partial p_1}{\partial t}+w\frac{\partial p_1}{\partial x}+c^2\rho div\vec{V_1}=0; v=\frac{4}{3}v_1+v_2$$

In (2.2) $v_1=\eta/\rho; v_2=\varsigma/\rho$ are the shear and volume kinematic viscosity, respectively. As in [8-10, 15], for simplicity, we neglect the influence of the acceleration due to gravity and other external forces.

In (2.2), the disturbance fields are indicated by the subscript 1, when in isentropic approximation the ratio between small density and pressure disturbances in this region has the form $p_1=c^2\rho_1$ [8-10]. Therefore, the equation for the pressure field disturbances in (2.2) follows from the continuity equation for the density disturbance.

Let us consider system (2.2) for the acoustic disturbances, without considering the evolution of entropy disturbances behind a shock. At zero viscosity system (2.2) exactly coincides with the discussed in [8, 9] system of equations for acoustic disturbances. As in [8, 9], we consider system (2.2) together with the boundary conditions on the surface of discontinuity determined by function (2.1), taking into account the fact that for the tangential and normal unit vectors to this surface we have the representations [9]:

$$\vec{t}=(t_x,t_z)=(-g_z,-1)/\sqrt{1+g_z^2};$$
$$\vec{n}=(n_x,n_z)=(1,-g_z)/\sqrt{1+g_z^2} \qquad (2.3)$$

Notation $g_z=\partial g/\partial z$ was introduced in (2.3).

From the condition of continuity in the velocity field tangential component on the disturbed front of the shock wave, follows the equality of scalar products with vector $\vec{t}$ for the velocity vectors to the left and right sides of the front $(w+V_{1x},V_{1z})\circ\vec{t}=(w_0,0)\circ\vec{t}$. In linear approximation $|g_z|<<1$ from (2.3) we obtain [9]:

$$V_{1z}=g_z(w_0-w) \qquad (2.4)$$

The boundary condition for the normal component of the velocity field is determined as the difference of the scalar product of vector $\vec{n}$ from (2.3) with the velocities on the left $x<0$ and right $x>0$ sides from the front in the form of $(w+V_{1x},V_{1z})\circ\vec{n}-(w_0,0)\circ\vec{n}=w-w_0+w_1$.

The values $w_0; w$ in (2.4) and in a boundary condition for normal velocity field components are determined from the equations of mass and momentum balance at the discontinuity, when it is taken into account the effect of viscosity [10]. As shown below, the destabilizing effect of viscosity occurs only when taking into account the corresponding terms in (2.2). By taking into account the viscosity in the momentum balance equation at the discontinuity leads only to a stabilization of the system in case of a sufficiently high viscosity. Therefore, a new instability condition (1.3) corresponds to the usual consideration of balance condition at the discontinuity in the limit of small viscosity.

In this regard, let us take the viscosity into account approximately in the known representation of the momentum balance equation (see (93.2) in [10]). We obtain a modification of this representation for the front having a finite small width $\lambda$ due to viscosity [13]. At the same time, as a result of approximate substitution $(\frac{4}{3}\eta+\varsigma)j\frac{dV}{dx}\cong(\frac{4}{3}\eta+\varsigma)|j|V/\lambda$ in formula (93.2) in [10], we obtain the quadratic equation for a constant mass flow $j$:

$$j^2+|j|\frac{\alpha\lambda\rho_0}{(\delta-1)}-\rho_0^2 F_0=0;$$

$$F_0=\frac{(p-p_0)\delta}{\rho_0(\delta-1)}; \delta=\frac{\rho}{\rho_0}; |j|=\rho_0|w_0|=\rho|w|; \alpha=\frac{v_0}{\lambda^2}, v_0=(\frac{4}{3}\eta+\varsigma)/\rho_0 \quad (2.5)$$

In (2.5) a coefficient of friction $\alpha$ is introduced which is determined by the value of the total coefficient of kinematic viscosity $v_0\equiv v\delta$ and front width $\lambda$.

Thus, the value of disturbance $w_1$ must be determined from representation for value $w-w_0$, which follows from the solution of the quadratic equation (2.5) for velocity $|w_0|$:

$$w-w_0=\frac{(\delta-1)}{\delta}|w_0|; w_0^2=F(\alpha); F(\alpha=0)=F_0$$

$$F(\alpha)=F_0-\frac{\alpha\lambda}{2(\delta-1)^2}(\sqrt{q}-\alpha\lambda); q=\alpha^2\lambda^2+4F_0(\delta-1)^2 \quad (2.6)$$

In this representation, it is necessary to make a substitution $p\to p+p_1; \rho\to\rho+\rho_1$ and expand it into a Taylor series within the limit $p_1\ll p; \rho_1\ll\rho$ in which the terms quadratic in the disturbance of density and pressure should be dropped. As a result, the boundary condition for the normal component of the velocity field leads to the equation which has the following form at value:

$$V_{1x} = \frac{(\delta-1)F_0}{2\delta\sqrt{F}}(\frac{p_1}{p-p_0}(1-\frac{\lambda\alpha}{\sqrt{q}})+\frac{\rho_1}{\rho_0\delta(\delta-1)}A);$$

$$A = \frac{2F}{F_0}-1-\lambda\alpha B; B = \frac{2\delta-1}{\sqrt{q}}-\frac{\delta(\sqrt{q}-\lambda\alpha)}{(\delta-1)^2 F_0} \qquad (2.7)$$

In case of zero friction $\alpha = 0$ equation (2.7) exactly coincides with the equation given in [9]. In (2.4) it is also necessary to take into account the solution of equation (2.5) in the form of (2.6). As in the theory of [8, 9], in (2.7) we use the relationship between disturbances of density and pressure on the Hugoniot shock adiabat in the following form:

$$p_1 = (\frac{dp}{d\rho})_H \rho_1 = -\rho^2(\frac{dp}{dV})_H \rho_1 \qquad (2.8)$$

To find the equation that determines the form of function (2.1), we use the equality that determines the disturbance of the shock wave velocity in the form $D_1 = \partial g/\partial t$, as well as the relation obtained for value $w_0^2 = (U_0-D)^2 = D^2$ in (2.6). At the same time $D \to D+D_1$ should be substituted in the left side of the expression (2.6) for the square of velocity $w_0^2$, and $p \to p+p_1; \rho \to \rho+\rho_1$ substituted in the right side and expanded into Taylor series. It is necessary to consider only the terms of not higher than the first order infinitesimal disturbances. Thus, we obtain the following equation:

$$\frac{\partial g}{\partial t} = \frac{F_0}{2\sqrt{F}}(\frac{p_1}{p-p_0}(1-\frac{\lambda\alpha}{\sqrt{q}})-\frac{\rho_1}{\rho_0\delta(\delta-1)}(1+\lambda\alpha B)) \qquad (2.9)$$

In case of zero friction equation (2.9) coincides with the equation given in [9].

As in [8, 9], we look for a solution to equations (2.2) under boundary conditions (2.4)-(2.9) in the following form:

$$x_{sh} = g(z,t) = \bar{g}\exp(i(kz-\omega t));$$

$$(V_{1x},V_{1z},p_1) = (\bar{V}_{1x},\bar{V}_{1z},\bar{p}_1)\exp(i(kz+lx-\omega t)) \qquad (2.10)$$

In this case, the fulfillment of the zero boundary condition at infinity far from front $x \to -\infty$ for the disturbances of velocity and pressure in (2.10) is required. This imposes a limitation on the magnitude of the longitudinal wave number $l$, the imaginary part of which must be non-zero and negative $\text{Im} \, l < 0$. The amplitude of the shock front disturbances in (2.10) is determined by the displacement of the front sections along the $x$ axis and these displacements are different depending on the coordinate $z$, which is characterized by the real wave number $k$. For zero wave number $k = 0$ the case of one-dimensional longitudinal disturbances may be obtained.

2. From the solvability condition of system (2.2), for solutions in the form of (2.10) we obtain the following dispersion equation:

$$\omega = -i\alpha_1 + wl \pm \sqrt{c^2(k^2 + l^2) - \alpha_1^2}; \alpha_1 = \frac{\nu}{2}(k^2 + l^2); \nu = \frac{4}{3}\nu_1 + \nu_2 \qquad (2.11)$$

The dispersion equation (2.11) with zero viscosity $\alpha_1 = 0$ exactly coincides with the equation given in [8, 9] (in [9] see (18)). In (2.11) it is necessary to select the case that corresponds to the minus sign that corresponds to the absence of sustained oscillations within the limit of high viscosity, when $\alpha_1^2 >> c^2(k^2 + l^2)$ in (2.11). Also in case of zero volume viscosity $\nu_2 = 0$ equation (2.11) exactly coincides with the dispersion equation given in [15] (see (32) in [15]) for disturbances of acoustic type.

3. In [15], system of equations (2.2) is considered in conjunction with the equation for the entropy $s_1$ disturbance, which in the context of representation (2.10) has the form of $(\omega - wl)\bar{s}_1 = 0$. At the same time, in case of non-zero entropy $\bar{s}_1 \neq 0$ disturbance, equality $l = l_e = \omega/w$ should be fulfilled. Besides, from (2.2), for disturbances of entropy type in [15], the following system of equations is obtained:

$$i\nu_1(\frac{4l_e^2}{3} + k^2)V_{ez} - \frac{1}{3}i\nu_1 k l_e V_{ex} + Vl\bar{p}_{1e} = 0;$$

$$-i\nu_1(l_e^2 + \frac{4}{3}k^2)V_{ex} - \frac{1}{3}i\nu_1 k l_e V_{ez} + Vk\bar{p}_{1e} = 0; \qquad (2.12)$$

$$kV_{ex} + l_e V_{ez} = 0$$

Multiplying the first equation in (2.12) by $l_e$ and the second one by $k$ and then adding the result, we obtain equation $(k^2 + l_e^2)(p_e V - i\frac{4}{3}\nu_1(kV_{ex} + l_e V_{ez})) = 0$. From the above equation, taking into account the third equation in (2.12), the equation $(k^2 + l_e^2)\bar{p}_{1e} = 0$ is obtained. Thus, the conclusion is made in [15] that the disturbance of entropy type for pressure is equal to zero $\bar{p}_{1e} = 0$. At the same time, in [15] it is implicitly assumed that $k^2 + l_s^2 \neq 0$.

This assumption, however, at a finite viscosity in (2.12) leads to a contradiction. Indeed, if in (2.12) the pressure is zero $\bar{p}_{1e} = 0$, the remaining terms in the first two equations (2.12) give a non-zero solution for velocity field disturbance of entropy type only under a condition $k^2 + l_e^2 = 0$ which is not taken into account in [15].

It indicates that, taking into account the viscosity, it is improperly to consider disturbances of entropy type in the adiabatic approximation, even in the linear formulation of the problem. From the other side, the disturbances of entropy do not determine the evolution of acoustic disturbances in (2.2) which may be considered independently from the entropy disturbances as it is provided in this paper.

4. Additionally to (2.11), the dispersion equation can be obtained from the boundary conditions (2.4) - (2.9) by taking into account the representation of the solution in the form of (2.10). At the same time, we can exclude the unknown function $g$ from (2.4) and (2.9). As a result, from (2.10), (2.7) and (2.8) the following system of equations is obtained:

$$\bar{V}_{1x} = \frac{a_2}{2\rho_0 \sqrt{F}} \bar{p}_1;$$

$$\bar{V}_{1z} = \frac{a_1 k}{2\omega \rho_0} \bar{p}_1; \qquad (2.13)$$

$$a_1 = 1 - \frac{\lambda \alpha}{\sqrt{q}} + \frac{hF_0}{F}(1 + \lambda \alpha B); a_2 = a_1 - 2h; h = j^2 (\frac{dV}{dp})_H$$

To close the system (2.13), we use an additional equation that determines the relationship between the pressure $\bar{p}_1$ and the velocity field $\bar{V}_{1x}; \bar{V}_{1z}$ following from (2.2) and (2.10) and having the following form:

$$\bar{p}_1 = \frac{c^2 \rho (l\bar{V}_{1x} + k\bar{V}_{1z})}{\omega - wl} \qquad (2.14)$$

From the solvability condition of system (2.13), (2.14) we obtain the dispersion equation additional to (2.11), which has the form given below:

$$\omega + clM(1 - \frac{a_2(h)}{2M^2}) = \frac{a_1(h)}{2} \frac{k^2 c^2 \delta}{\omega} \qquad (2.15)$$

In (2.13) and (2.15) $h$ is the D'yakov parameter. For functions $a_1(h), a_2(h)$ within the limit $\lambda \alpha \ll \sqrt{F_0}$ we have the following representations:

$$a_1 = 1 + h - \frac{\lambda\alpha}{2\sqrt{F_0(\delta-1)}}(1+h(2\delta-1)) + O(\frac{\alpha^2\lambda^2}{F_0}); a_2 = a_1 - 2h \quad (2.16)$$

In the opposite limit $\varepsilon = \frac{F_0(\delta-1)}{\lambda^2\alpha^2} \ll 1$:

$$a_1 = -2h(\delta-1) + 2\varepsilon(1+h(2\delta-1)) + O(\varepsilon^2); a_2 = a_1 - 2h \quad (2.17)$$

Note that limit $\varepsilon \ll 1$ and estimate in (2.17) correspond to the case of small Reynolds numbers defined by the finite front width $\sqrt{\varepsilon} \approx \text{Re}_\lambda = \frac{|w_0|\lambda}{\nu} \ll 1$.

### 3. Conditions of the shock wave instability

In general, in case of finite viscosity in (2.11) and (2.15) we can use wave number representation in the form of $k = \frac{\Omega}{c}\cos\theta; l = \frac{\Omega}{c}\sin\theta$ [8-10]. If we accept that $\sin\theta = -i\rho; \cos\theta = \sqrt{1+\rho^2}; \rho > 0; \text{Im}\,\rho = 0$, then for the longitudinal wave number we obtain:

$$l = -i\frac{k\rho}{\sqrt{1+\rho^2}} \quad (3.1)$$

From (2.11), for the frequency we have the following:

$$\omega = \Omega(F - M\sin\theta) = \frac{kc}{\sqrt{1+\rho^2}}(iM\rho + F);$$

$$F = -i\frac{\nu k}{2c\sqrt{1+\rho^2}} - \sqrt{1 - \frac{\nu^2 k^2}{4c^2(1+\rho^2)}}. \quad (3.2)$$

The shock wave instability can occur when in (3.1) and (3.2) conditions $\text{Im}\,l < 0; \text{Im}\,\omega > 0$ are simultaneously satisfied [9]. Inequality $\text{Im}\,l < 0$ is present in any positive real values $\rho$, and $\text{Im}\,\omega > 0$ is performed only at sufficiently high Reynolds numbers:

$$\text{Re}_k = \frac{2|w|}{k\nu} > \text{Re}_{th};$$

$$\text{Re}_{th} = \frac{1}{\rho\sqrt{1+\rho^2}}; \frac{k\nu}{2c\sqrt{1+\rho^2}} < 1; \quad (3.3)$$

$$\text{Re}_{th} = \frac{1}{\rho\sqrt{1+\rho^2}}(1 + \sqrt{1 - \frac{4c^2(1+\rho^2)}{\nu^2 k^2}}); \frac{k\nu}{2c\sqrt{1+\rho^2}} > 1$$

Value $\rho$, appeared in (3.1) - (3.3), must be determined from the real and positive solution of the equation following from (2.15), (3.1), (3.2) and having form:

$$\rho^2 + i\rho F(\rho)\frac{(a_2 + 2M^2)}{M(a_2 + a_1\delta)} + \frac{a_1\delta - 2F^2(\rho)}{a_2 + a_1\delta} = 0 \qquad (3.4)$$

At zero viscosity $F(\rho) = -1$ in (3.4) and this equation has no solution with $\operatorname{Im}\rho = 0$. In case of finite viscosity equation (3.4) solution with $\operatorname{Im}\rho = 0$ can exist, but only at the following condition (see (3.3)):

$$\frac{1}{1+\rho^2} > \operatorname{Re}_S^2 > \frac{\rho(2-\rho)}{1+\rho^2};$$

$$\operatorname{Re}_S = \frac{2c}{k\nu}; \rho > 1 \qquad (3.5)$$

Under condition (3.5) value $F$ in (3.2) and (3.4) already has a zero, real part, and, at the same time, equation (3.4) can have a positive real solution with a zero imaginary part.

Note that in case of finite viscosity equation (3.4) is not a quadratic equation for unknown value $\rho$ any more, as in [8,9]. Therefore, in general, it is difficult to obtain analytical expressions for the conditions of instability and, in this regard, the following sections consider two particular cases related to the limits of the long $k \to 0$ and short $k^2 >> |l^2|$ waves of transversal disturbances. Besides, the first case corresponds to the case of one-dimensional longitudinal disturbances.

### 4. Instability to the longitudinal one-dimensional disturbances.

Let us consider the possibility of shock instability with respect to the longitudinal one-dimensional disturbances when in (2.11) and (2.15) the limit of zero transverse wave number $k \to 0$ takes place. In this case, from (2.11) and (2.15) for the longitudinal wave number and frequency we obtain the solutions in the following form:

$$l = -i\frac{2cM}{a_2 v}(1 - \frac{a_2^2}{4M^2}) \qquad (4.1)$$

$$\omega = -lcM(1 - \frac{a_2}{2M^2}) = i\frac{2w^2}{a_2 v}(1 - \frac{a_2^2}{4M^2})(1 - \frac{a_2}{2M^2}) \qquad (4.2)$$

For instability of the shock wave inequalities $\mathrm{Im}\, l < 0; \mathrm{Im}\, \omega > 0$ should be fulfilled [9]. Thus, it is possible if in (4.1) and (4.2) fulfilled is either case a) $0 < a_2 < 2M^2$ or case b) $a_2 = -|a_2| < 0; |a_2| > 2M$.

Within the limit of small value of friction $\alpha \to 0$, taking into account (2.16), the following condition for the Dyakov parameter, at which the instability occurs, corresponds to case a):

$$1 - 2M^2(0) + \frac{\lambda \alpha M^2(0)(2\delta+1)}{\sqrt{F_0(\delta-1)}}(1 - \frac{M_{th}^2}{M^2(0)}) < h < 1 - \frac{\lambda \alpha \delta}{\sqrt{F_0(\delta-1)}} + O(\frac{\lambda^2 \alpha^2}{F_0});$$

$$M(0) = M(\alpha = 0); M_{th} = \sqrt{\frac{\delta}{2\delta+1}}$$

(4.3)

For zero friction in (2.6) from (4.3) follows an absolute instability condition (1.3), where $M = M(0)$.

Within the same limit the following condition of instability corresponds to case b):

$$h > (1 + 2M(0))(1 - \frac{\lambda \alpha \delta}{\sqrt{F_0(\delta-1)}}) + O(\frac{\lambda^2 \alpha^2}{F_0})$$

(4.4)

Note that in case of zero friction $\alpha = 0$ condition of instability (4.4) coincides with condition of instability (1.1).

In the opposite limit of relatively large uniform friction for case a) the instability is impossible, and in case b) taking into account (2.17) we obtain the condition of instability in the following form:

$$h > \frac{2M(0)}{\delta}\sqrt{\varepsilon} + O(\varepsilon)$$

(4.5)

Thus, all the obtained conditions of the shock wave plane front instability (4.3) - (4.5) are corresponded to the same representation for the increment of exponential growth of one-dimensional longitudinal disturbances which is defined by positive imaginary part of the frequency in (4.2). The increment quantity, as is easily seen from (4.2), is proportional to $\tau^{-1} = \frac{2w^2}{v}$ quantity having a dimension of inverse time. The existence of parameter with dimension of inverse time is need as necessary (but not sufficient, see [12]) condition for instability realizing [10].

## 5. Corrugation instability

Let us consider the dispersion equations (2.11) and (2.15) for the case of two-dimensional (corrugation) disturbances in the opposite relatively short wavelength limit $k^2 >> |l^2|$ for the transversal two-dimensional disturbances. In this case, from (2.11) for the longitudinal wave number we obtain the expression in the following form:

$$l = -\frac{\omega + K}{|w|}; K = ik^2 \nu (1 + \sqrt{1 - \text{Re}_S^2})/2; \text{Re}_S = \frac{2c}{k\nu} \tag{5.1}$$

By taken into account (5.1) from (2.15) the quadratic equation for frequency is obtained, the solution of which has the following form:

$$\omega = \frac{K(2M^2 - a_2)}{2a_2}\left[1 \pm \sqrt{1 + \frac{4k^2 w^2 a_1 a_2 \delta}{K^2(2M^2 - a_2)^2}}\right] \tag{5.2}$$

From (5.1) and (5.2) it follows that the conditions of shock instability with respect to two-dimensional disturbances in case when $\text{Im}\,\omega > 0; \text{Im}\,l < 0$ are fulfilled when inequality $0 < a_2 < 2M^2$ takes place, when $\text{Re}_S < 1$ in (5.1). This is exactly the same as with the case of a), discussed above for one-dimensional disturbances. Therefore, for the case of two-dimensional corrugation disturbances the instability also occurs only when conditions (4.3) or (1.3) in the limit of $\alpha \to 0$ are fulfilled. The case b), considered for one-dimensional disturbances, according to (5.2), do not give possibility for instability at all.

Thus, it is considered the case $\text{Re}_S < 1$ in which the real parts of frequency (5.2) and longitudinal wave number (5.1) are equal to zero and absolute instability of shock realizes only in monotonic exponential increasing of disturbances amplitudes.

However, it follows from (5.1) and (5.2) that for Reynolds numbers $\text{Re}_S > 1$ an implementation of oscillatory instability is also possible when the real part of frequency in (5.2) may be nonzero.

From (5.2) it follows that the value of the exponential growth increment in corrugation two-dimensional disturbances is equal to $\tau_c^{-1} = k^2 \nu / 2$ and in the limit of low viscosity when inequality $\tau_c^{-1} << \tau^{-1} = \frac{2w^2}{\nu}$ is fulfilled it is much smaller than the increment in one-dimensional disturbances growth. Therefore, the instability with respect to small longitudinal one-dimensional disturbances can be dominant at the linear stage of disturbances evolution. At the same time it is allowed to assume that only at the nonlinear stage of development of one-dimensional disturbances the two-dimensional corrugation disturbances, observed in experiment [11], can already manifest themselves.

## 6. Comparison to the experiment and numerical simulation

We use the instability condition (1.3) (followed from condition (4.3)) for examples, when the value of the D'yakov parameter $h$ can be determined from experimental data.

6.1 Shock adiabat for real gas and fluid.

To find the analytical form of the D'yakov parameter $h$ we use the known from experimental data, a linear relationship between the shock wave velocity and the velocity of medium in the compression region behind the shock wave front [17-19]:

$$D = A + BU \qquad (6.1)$$

In (6.1), it is assumed that quantities $A, B$ can be regarded as constants within some range of changes in the compression quantity $\delta = 1/y = V_0/V$.

For example, in the case of shocks in water within a range of changes in the velocity behind the front $7.1 km/s > U > 1.5 km/s$ in (6.1) $A = 2.393 km/s; B = 1.333$ [18].

Also, from the observation data on shock waves in air [23] it follows that within the range of velocities $3.982 km/s > U > 1.705 km/s$ in (6.1) values $A = 0.215 km/s; B = 1.0597$ should be presented. From observations on shock waves in argon [24] for the velocity range $7.81 km/s > U > 3.03 km/s$ of value $A = 0.819 km/s; B = 1.009$ and for shock waves in nitrogen [25] $A = 0.386 km/s; B = 1.04046$ when $8.99 km/s > U > 3.8 km/s$.

On the basis of (6.1) and taking into account the Rankine-Hugoniot relations, recorded in the form of $x = \frac{p}{p_0} = 1 + DU/p_0 V_0; \frac{1}{\delta} = y = V/V_0 = 1 - \frac{U}{D}$, the following representation for Hugoniot curve $x = 1 + \frac{\rho_0 A^2 (1-y)}{p_0 (1 - B + By)^2}$ is obtained. Also the Mach number in the compression region $M = \frac{A}{(B - \delta(B-1))c}$ and the D'yakov parameter $h = -\frac{B - \delta(B-1)}{(B+1)\delta - B}$ representations are obtained. In this case, for values of compression $1 < \delta < \delta_{max} = \frac{B}{B-1}$ the D'yakov parameter can take only negative values $h < 0$. But formal it is possible consider also the case when $\delta > \delta_{max}$ and $h > 0$. Only for the case $\delta < \delta_{max}$ our consideration is similar with the case of the polytropic gas (with polytropic index $\gamma > 1$), for which $h = -\dfrac{1}{M_0^2}; \delta_{max} = \dfrac{\gamma+1}{\gamma-1}$ [10].

Thus, from (6.1) and (1.3) the condition of shock absolute instability is obtained in the form:

$$\frac{c^2}{A^2} < \frac{(B+1)\delta - B}{\delta(B - \delta(B-1))^2} \qquad (6.2)$$

In (6.2) it is still required to determine the dependence of sound velocity in the compression region $c$ from values of compression $\delta$. In order to do it, in the case of shock waves in gas, we may use known ratio $c^2 = p\gamma/\rho = x(y)c_0^2/\delta$, which leads to the following representation:

$$c^2 = \frac{c_0^2}{\delta}\left[1 + \frac{\delta(\delta-1)A^2\gamma}{c_0^2(B-\delta(B-1))^2}\right] \qquad (6.3)$$

From (6.2) and (6.3) the condition for the linear absolute instability is obtained in the form:

$$1 < \delta_- < \delta = \frac{DB}{A+D(B-1)} < \delta_{\div};$$

$$\delta_{\pm} = a \pm \sqrt{a^2 - b}; \qquad (6.4)$$

$$a = \frac{B(B-1) + \dfrac{A^2}{2c_0^2}(B+1+\gamma)}{(B-1)^2 + \dfrac{A^2\gamma}{c_0^2}}; b = \frac{B^2 + \dfrac{A^2}{c_0^2}}{(B-1)^2 + \dfrac{A^2\gamma}{c_0^2}}$$

From (6.4) it follows that the shock wave front instability in gas can be realized only when setting a lower and upper bound on the magnitude of the shock wave velocity:

$$D_{min} = D_- < D < D_{max} = D_{\div};$$

$$D_{\pm} = \frac{A\delta_{\pm}}{B - \delta_{\pm}(B-1)} \qquad (6.5)$$

Thus, for realization of the shock wave plane front instability in a real gas condition (6.5) must be fulfilled if (6.1) is take place.

Indeed, in experiment [11] it was found that in argon the instability of plane shock wave is observed only when the shock wave velocities belong to limited, narrow interval $11.5 km/s > D > 10 km/s$ and in carbon dioxide only to velocity range $6 km/s > D > 5 km/s$, but only for the positive value $h \approx 0.05$.

Besides, in [11] for a shock wave in carbon dioxide the instability is also observed at low negative values of the D'yakov parameter near value $h \approx -0.02$, when the shock wave velocity is equal to $D \approx 3.5 km/s$. This qualitatively corresponds to the case when under conditions (6.4) and (6.5) the velocity interval degenerates to a single point.

6.2 Instability of shock waves in fluids.

The condition of the shock wave instability in a fluid, as well as for gases, is defined by inequality (6.2). In contrast to the shock waves in gases in (6.2), instead of representation (6.3), for the square of sound velocity in the compression region, another expression must be used [26, 27]:

$$c^2 = \frac{dp}{d\rho} = c_0^2 \delta^{n-1};$$

$$p - p_0 = \frac{\rho_0 c_0^2}{n}(\delta^n - 1) \qquad (6.6)$$

According to [27], in (6.6) exponent $n = 7.15$ can be used for pressure up to 25 kbar. In [17] for shock waves in water given is data on sound velocity in the compression region, from which it follows that by increasing the pressure a tendency on decreasing this parameter takes place. For example, $n = 6.14$ when pressure is $p = 70 kbar$ and $n = 5.26$ when pressure is $p = 250 kbar$. If $A = c_0 = 1.483 km/s; B = 2.118$ in (6.1) [17-19], then from the instability condition (6.2) it follows that the plane shock wave front in water becomes unstable when:

$$\delta_{min} = 1.5617 < \delta < \delta_{max} = B/(B-1) = 1.894 \qquad (6.7)$$

Besides, for a lower bound of the compression value (6.7) it is assumed that this threshold compression value corresponds to the experimental data [17]. According to the given data, when using the indicated in (6.7) threshold compression $p = 70 kbar$ $D = 4.414 km/s, U = 1.588 km/s$ and the sound velocity is $c = 4.666 km/s$ that corresponds to $n = 6.14$ in (6.6). As a result, we obtain that the instability of plane shock waves in water may take place in the range of compression values (6.7), which according to the experimental data [17] corresponds to a range of the shock wave velocity in the following form:

$$D_{min} = 4.414 km/s < D < D_{max} = 6.679 km/s, \qquad (6.8)$$

Such shock wave velocities have been observed in [17] in the pressure range $70 kbar \leq p \leq 210 kbar$.

6.3 Comparison to the numerical simulation results

In [20] the results of numerical study of shock waves stability are given (see chap. 12 in [20] and [21, 22]). It is shown that on the nonlinear stage the increasing of disturbances amplitude which occurs in a situation when in classical theory [8-10] only neutral stability (with constant amplitude of oscillations) is realizing under the following condition:

$$h_{min} = \frac{1 - M^2(1+\delta)}{1+(\delta-1)M^2} < h < h_{max} = 1 + 2M \qquad (6.9)$$

Furthermore, it is noted in [20] that when fulfilling condition (1.1) an instability associated with forming a structure of two shock waves following each other also takes place. At the same time for the D'yakov parameter outside conditions (6.9) or (1.1) the shock wave was quite stable [20] when $-1 < h < h_{min}$.

It follows from a comparison of (6.9) and obtained in the present study instability condition (1.3) that the interval of the D'yakov parameter variation defined in (1.3) completely lays inside of the interval for that parameter variation under condition (6.9). Indeed, an upper bound in (1.3) is strictly lower than maximum value in (6.9) because $1 < h_{max}$ for any positive $M > 0$. So the minimum value in (1.3) is larger than minimum value in (6.9) because $h_{min} < 1 - 2M^2$ for any $0 < M < 1$ when $\delta > 1$. Note that condition (1.3) corresponds to realization of absolute instability of plane shock wave front obtained within the framework of the linear stability theory. At the same time it is concordant with the above results of numerical simulation of nonlinear evolution in finite amplitude disturbances.

Since the numerical simulation inevitably requires entering the numerical viscosity it is not surprising that there is a correspondence between the obtained in the present study instability condition (1.3) and the results of numerical simulation of a plane shock wave front instability in [20].

The condition of shock wave instability (1.3) is also concordant with the results of numerical simulation [7] of shock waves formed in supernova explosions, when estimate $-10^{-3} \leq h \leq 10^{-3}$ is given for the Dyakov parameter (the author is grateful to D.A. Badzhin for providing relevant estmates). Indeed, in [7] $D \approx 100 km/s; 4 < \delta < 6000, c \approx 0.83 km/s$ and condition (1.3) turn out to be fulfilled, for

417example, for value $h \approx -10^{-3}$ in case when values of Mach number and compression satisfy conditions $1 > M \geq 0.71; \delta \leq 170$.

By taking into account (4.1) and (4.2) it is possible to estimate the characteristic time of exponential increase of acoustic disturbances in shock waves in water $\tau \cong (2w^2/\nu)^{-1} \approx 10^{-12} \sec$, assuming that $\nu \approx O(10^{-6} m^2/\sec); |w| \approx O(10^3 m/\sec)$ [2]. At the same time the characteristic scale of changing the disturbances in space is $l^{-1} = L_x \approx 10^{-9} m$.

For astrophysical scales corresponding to numerical simulation in [7] for average coefficient of kinematic viscosity (in the transient layer) $\nu \cong 4.5 \times 10^8 km^2/s$ and magnitude of the velocity jump on shock $|w| = 0.6 km/s$ it is possible to obtain estimation $\tau \cong 6.25 \times 10^8 \sec \approx 19.9 years$. However, in case of taking into account the magnitude of the numerical (circuit) viscosity $\nu_{num} \approx 5 \times 10^{11} km^2/s$, this estimate is much higher $\tau \cong 6.945 \times 10^{11} \sec \approx 22100 years$. Because the typical time of the shock wave existence is near $2 \times 10^5 years$, in the framework of a numerical model [7], this time substantially exceeds the mentioned estimation. That indicates the admissibility of realization of the considered herein dissipative mechanism of shock wave instability in model [7].

It follows from the above estimates that the considered acoustic disturbances decrease rapidly with distance from the shock wave front. They become vanishingly small at distances commensurable with the width of the shock wave front (in [7], the width of the transient region $\lambda \cong 3.6 \times 10^{11} km$). Dissipative instability, implemented primarily in flow behind the shock wave front by type of instability in the boundary layer, leads to instability of the front itself observed in the experiment.

## Conclusions

New conditions of instability of the plane shock wave in arbitrary media, by taking into account the media viscosity, are obtained. This has been achieved by considering only the acoustic type disturbances. A condition of instability with respect to one- and two-dimensional disturbances, which has the same form for these two types of disturbances in the limit of low viscosity is obtained. At the same time the increment of instability with respect to one-dimensional longitudinal disturbances in the given limit significantly exceeds the increment of the exponential growth in time of usually considered corrugation two-dimensional disturbances. This means that on the linear stage of instability just the longitudinal one-dimensional disturbances are important and only on the nonlinear stage they can lead to the observable in [11] corrugation instability.

In the present paper a new mechanism of dissipative instability for plane shock wave front is discovered. The physical meaning of this mechanism is the implementation of flow instability behind the front at sufficiently high Reynolds numbers that is similar to instabilities in the boundary layer.

The given example of instability broadens the range of physical problems where a similar phenomenon of dissipative instability was studied earlier and led to a new comprehension of the corresponding phenomena basic mechanisms [16, 28, 29]. Considered is the application of the theory conclusions when the instability region (6.2) derived from condition (1.3) for a shock wave in real gases and fluids is obtained on the basis of the shock adiabat explicit form (6.1). It is important that the representation for the shock adiabat (6.1) is obtained directly from the experimental data without involving any model of the equation of state for a given medium. It is shown in [30] that the use of the pressure representations in the Navier–Stokes equation for a compressible medium based on the assumption of local thermodynamic equilibrium can lead to incorrect conclusions. Besides, the conclusions obtained in [30] can be used in connection with the reviewed in [31, 32] analogy between the instability of shock wave and numerical instability in solving Euler equations. From the estimate of the disturbances exponential growth increment it follows that the plane shock wave front instability is more implementable for the case of fluids compared with gases having substantially higher values of kinematic viscosity coefficients. Obtained are limitations (6.7) on the compression value in water, from which, in accordance with the experimental data [17], one can expect occurrence of plane shock wave front instability in water in the range of the shock wave velocities (6.8). However, from the obtained in the present work results follows a need for further development of theoretical and experimental studies of the problem of stability in shock waves, which still cannot be considered completely solved. Note that these results are obtained by generalizing the classical approach [8-10] to the investigation of shock waves stability by taking into account the viscosity. At the same time, not considered are additional possibilities which are given by, e.g., the approach proposed in [33], when considering the stability of the shock wave in a non-inertial coordinate system.

### Acknowledgements

I express gratitude to V.T. Gurovich and Ya.E. Krasik for support and attention at all stages of the paper preparation, as well as V.E. Fortov, A. Pirozhkov, A. Beloborodov, D. Badjin, M. Garasev and E. Kurbatov for their interest in the work results and our discussions at the FNP2019 conference.

The study is supported by Russian Science Foundation grant No.14-00806P and Israel Science Foundation grant No. 492/18.